# Continuous subsurface property retrieval from sparse radar observations using physics-informed neural networks

Ishfaq Aziz*[1], Mohamad Alipour[1]

[1] Department of Civil and Environmental Engineering, University of Illinois Urbana-Champaign.

## Abstract

Estimating subsurface dielectric properties is essential for applications ranging from environmental surveys of soils to nondestructive evaluation of concrete in infrastructure. Conventional wave-inversion methods typically assume few discrete homogeneous layers and require dense measurements or strong prior knowledge of material boundaries, limiting scalability and accuracy in realistic settings where properties vary continuously. We present a physics-informed machine learning framework that reconstructs subsurface permittivity as a fully neural, continuous function of depth, trained to satisfy both measurement data and Maxwell's equations. We validate the framework with both simulations and custom-built radar experiments on multi-layered natural materials. Results show close agreement with in-situ permittivity measurements ($R^2 \approx 0.93$), with sensitivity to even subtle variations ($\Delta \varepsilon_r \approx 2$). Parametric analysis reveals that accurate profiles can be recovered with as few as three strategically placed sensors in two-layer systems. This approach reframes subsurface inversion from boundary-driven to continuous property estimation, enabling accurate characterization of smooth permittivity variations and advancing electromagnetic imaging using low-cost radar systems.

______________________________________________________________________

## 1. Introduction

Estimating subsurface material properties is critical in diverse applications, including soil moisture estimation [1][2][3][4], permafrost characterization [5], quality evaluation in construction [6], biomass characterization for wildfire assessment [7], and monitoring building materials and infrastructure condition [8][9]. For instance, accurate estimation of soil and vegetation moisture enables effective irrigation management and wildfire risk assessment, resulting in substantial cost savings [10][11][12]. In civil infrastructure, moisture levels in subsurface concrete or asphalt are key indicators of deterioration in bridges and pavements, while the density of subsurface pavement layers informs pavement design and quality control.

Radar sensing has become a widely adopted tool for subsurface investigation in these domains because of its nondestructive nature, rapid data acquisition, and higher spatial coverage. In comparison, traditional methods such as using in situ sensors (e.g., time-domain reflectometry [13] and capacitance sensors [14]) and gravimetric oven drying – while accurate, are labor-intensive, time-consuming, and limited to point-scale measurements. Electromagnetic (EM) waves transmitted by radar (Fig. 1-a) are affected by the materials' dielectric properties, such as the dielectric constant or relative permittivity ($\varepsilon_r$) (Fig. 1-b). Relative permittivity of a material, defined as the ratio of its permittivity to that of vacuum ($\varepsilon_r = \varepsilon/\varepsilon_0$), will hereafter be referred to simply as 'permittivity.' Materials with higher permittivity reduce the EM wave velocity, delaying the arrival of received EM waves at the radar receiver (Fig. 1-b). These relative changes in received radar signals due to a change in permittivity can be leveraged to predict permittivity from radar signals.

A wide range of subsurface evaluations can be carried out based on permittivity, as changes in material characteristics, such as moisture, density, and defects, directly alter dielectric response. Notably, permittivity is particularly sensitive to moisture content, since water exhibits a very high permittivity ($\approx 81$) compared to air ($\approx 1$) or other materials, such as dry soil ($\approx 3-5$) and concrete ($\approx 5-9$). As a result, it has been widely used to estimate the moisture condition in soil [1][2], biomass [7], concrete [15], and asphalt materials [16]. In pavement evaluation, dielectric permittivity also serves as a key parameter in estimating the density of asphalt concrete [16].

Subsurface permittivity variation may be layered (Fig. 1-c), as in constructed pavements, or continuously varying (Fig. 1-d), as in smooth soil moisture gradients induced by rainfall, drought, or fire. Estimating these variations from received radar signals is an inverse problem, since the goal is to recover material properties from the measured wavefield. Traditional inversion methods, including full-waveform inversion [17][18][19], numerical model updating [20], and machine learning (ML) [7][21] are often tailored to discrete, layered settings and struggle to resolve gradual, continuous changes in permittivity. Wave inversion or model updating approaches typically assume a fixed number of discrete homogeneous layers with known interfaces. As the number of layers increases, the number of unknowns grows rapidly, leading to unconstrained parameter estimation and high uncertainty [20]. ML-based retrievals offer speed and automation, but require large volumes of training data, which are often expensive to collect. They also often generalize poorly to new scenarios.

Physics-informed neural networks (PINNs) offer an alternative that incorporates governing physics into the learning process and can be trained with very limited data [22][23][24][25]. Prior studies have successfully employed PINNs across various domains [24][26][27][28][29][30], though mostly with synthetic datasets. The incorporation of physics acts as a powerful form of regularization, enhancing generalization compared to purely data-driven ML methods and enabling learning from sparse measurements.

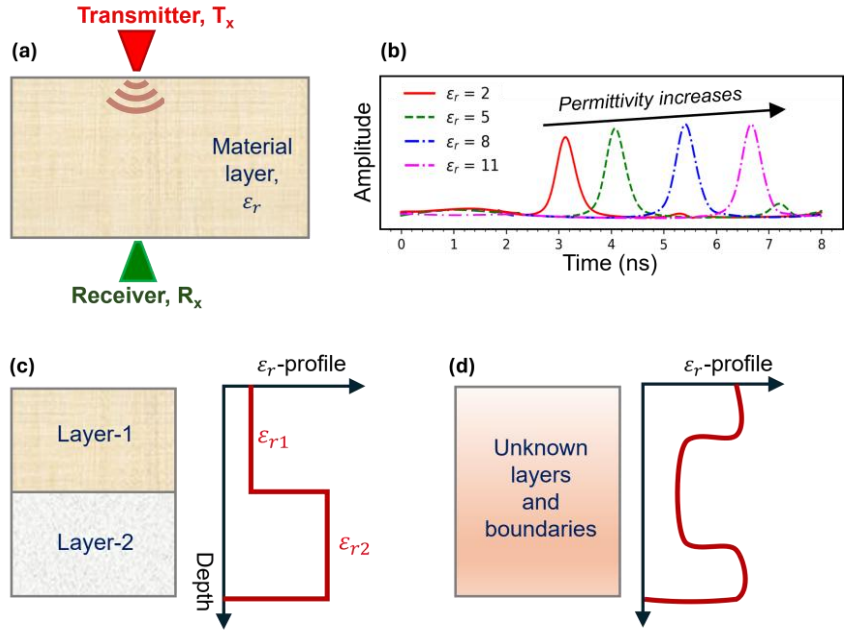

**Fig. 1: Radar sensing, concept of permittivity prediction in layered and continuous subsurface conditions.** (a) Radar transmitter-Tx sends EM wave through a material layer and receiver-Rx receives the transmitted wave; (b) Effect of change in received radar wave/signal due to a change in permittivity; (c) Multi-layer material with known discrete number of layers and known boundaries, and the corresponding $\varepsilon_r$-profile; (d) Unknown number of material layers and unknown boundaries, and corresponding continuous $\varepsilon_r$-profile retrieved.

We introduce a PINN-based inversion framework that reconstructs subsurface permittivity from sparse radar measurements, constrained by Maxwell's equations. Unlike traditional methods that assume a fixed number of discrete homogeneous layers, our approach represents permittivity as a fully neural, continuous function of depth, trained end-to-end to satisfy both measurement data and physics. This eliminates the need for prior assumptions about material boundaries or layer counts, and yields physically consistent, high-resolution reconstructions that capture gradual permittivity variations. In contrast to conventional electromagnetic inversion approaches, which require high-dimensional optimization, strong priors, and dense, noise-free data to remain stable [20][31], our method leverages the expressive power of neural networks to encode a continuous permittivity field, while enforcing Maxwell's equations to guide learning from only a handful of measured wavefield snapshots. This formulation overcomes longstanding limitations in multilayer permittivity estimation and enables a new class of physics-informed, data-efficient electromagnetic imaging techniques with broad implications for geophysical exploration, nondestructive testing, and environmental sensing.

The proposed method was applied to predict $\varepsilon_r$ in two scenarios: (i) a simplified setting with known layers and boundaries, similar to Fig. 1-c, and (ii) a more complex setting where both the number of layers and their boundary locations were unknown (Fig. 1-d). In scenario-i, the number of retrieved $\varepsilon_r$-values matches the number of known discrete layers, while in scenario-ii, the retrieved $\varepsilon_r$ will be a continuous profile along space (or depth).

We first implement the method on numerically generated synthetic data to validate its accuracy under known controlled conditions. Once validated, we applied it to waveforms collected from real experimental setups to retrieve the continuous spatial distribution of permittivity of real-world materials. We used the finite difference time domain (FDTD) method to generate synthetic data for training and testing the PINN framework.

For validation on real-world experimental data, we built a multi-layered material setup in the lab. We also introduce a custom radar setup for the PINN framework using a vector network analyzer (VNA) and radio frequency (RF) antennas. The custom radar setup is an order of magnitude cheaper than commercially available ground penetrating radars (GPR) and offers enhanced flexibility with capabilities of multi-frequency sensing and multi-receiver data collection.

We also conducted a parametric analysis to evaluate how the number and spatial placement of RF antennas affect the accuracy of the prediction of the permittivity profile, particularly in cases where there is a sharp contrast in permittivity across the interface of material layers.

## 2. Results and discussions

### 2.1. Prediction of material permittivity using physics-informed neural network (PINN)

To retrieve subsurface permittivity, we developed a framework integrating radar sensing and Physics-Informed Neural Network (PINN) models, where PINN incorporates sparse wave measurements with the governing equations of EM wave propagation. The process (shown in Fig. 2) begins with radar wave propagation through a multi-layered medium (Fig. 2-a), where an EM pulse emitted from a source interacts with different layers and is recorded by multiple spatially distributed receiver sensors (e.g., Rx-sensors in Fig. 2- a, b). These sensors capture waveform responses that encode information about the permittivity or dielectric constants and material boundaries. As illustrated in Fig. 2(b), the recorded waveforms vary depending on the location of the sensors and the underlying material distribution, as well as how the signal is altered during propagation, including delayed arrival of peaks, attenuation of amplitude, and multipath reflections within the medium. A PINN model can be trained with the data recorded by Rx-sensors to retrieve discrete permittivity values or a continuous permittivity profile of the layers (Fig. 2-c).

As shown in Fig. 2, we considered two options for the PINN model. The first option, PINN model-1 (Fig. 2-d), can be used when the number of discrete material layers and the spatial location of their boundaries are known. A single neural network is used here that predicts the electric field ($E$) with distance/space ($x$) and time ($t$) as inputs. The permittivity of each discrete layer ($\varepsilon_{r_i}$) was set as a learnable network parameter that is learned during training.

The second option, the PINN model-2 (shown in Fig. 2-e), can be used when the number of layers and their boundaries are unknown. Here, two networks were used − the field network and the permittivity network. The field network is the same as Fig. 2(d), except that $\varepsilon_r$ is not a learnable parameter. Instead, the permittivity network predicts $\varepsilon_r$ with distance/space ($x$) as input. Time is not an input to the permittivity network, as the permittivity is independent of time. The remaining details of the PINN are explained in the Methods section.

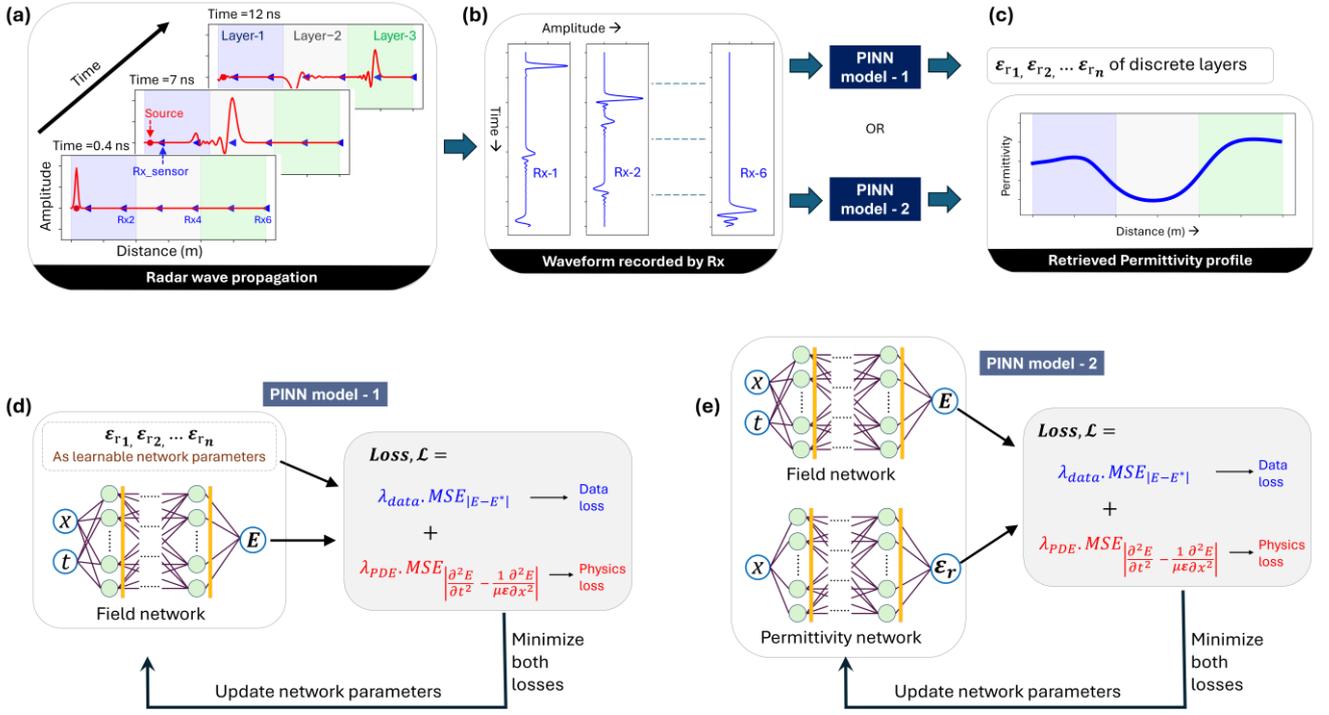

**Fig. 2: Outline of the proposed framework.** (a) Radar wave propagates in multi-layered media, and Rx-sensors are spatially distributed to record the waveform; (b) Waveforms recorded by spatially distributed Rx-sensors; (c) Permittivity retrieved by PINN model 1 and 2; (d) PINN model-1: Permittivity ($\varepsilon_{r_i}$) of discrete material layers are predicted as learnable parameters; (e) PINN model-2: Continuous permittivity profile is predicted with a separate network as a function of space.

## 2.2. Validation on synthetic data

To generate synthetic data, we conducted finite-difference-time-domain (FDTD) simulations for two-layer and three-layer material configurations (shown in Fig. 3), where the source emits a 1D electromagnetic wave and the receivers record it at various spatial locations. Details of FDTD simulations are described in the 'Methods' section. The PINN models were trained with the recorded waveform data, and the total loss with each epoch of training is depicted in Fig. 4(a). The loss gradually decreases with training, indicating proper training of the model.

Assuming the number of discrete material layers and the spatial location of the layer boundaries to be known, we used the PINN model-1 (Fig. 2-d) to retrieve the permittivity of these discrete layers. Fig. 4(b) shows the retrieved permittivity values against the true values for all different cases, depicting a good correlation of prediction. The predictions on the two and three-layer setups are shown with different labels.

Next, assuming the number of discrete material layers and the spatial location of the layer boundaries to be unknown, we used the PINN model-2 (Fig. 2-e) to retrieve continuous permittivity profiles along space. Fig. 4(c) shows the retrieved permittivity profiles for four different cases. The predicted profiles align well with the true profiles for both two-layer and three-layer material scenarios. Since a neural network is a continuous function, the permittivity profile predicted by PINN is smooth and does not produce abrupt changes at the layer boundaries.

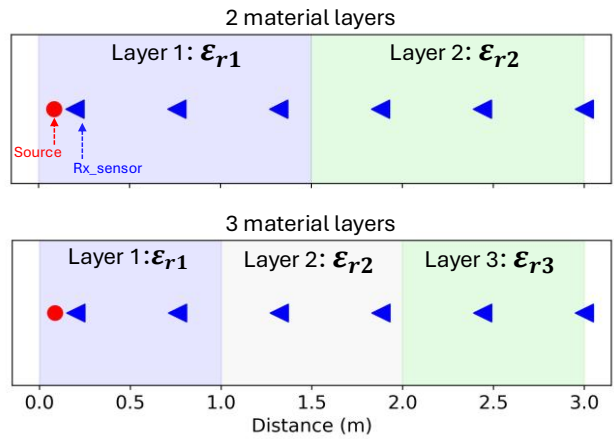

**Fig. 3: Two and three-material layer configurations in numerical simulations.**

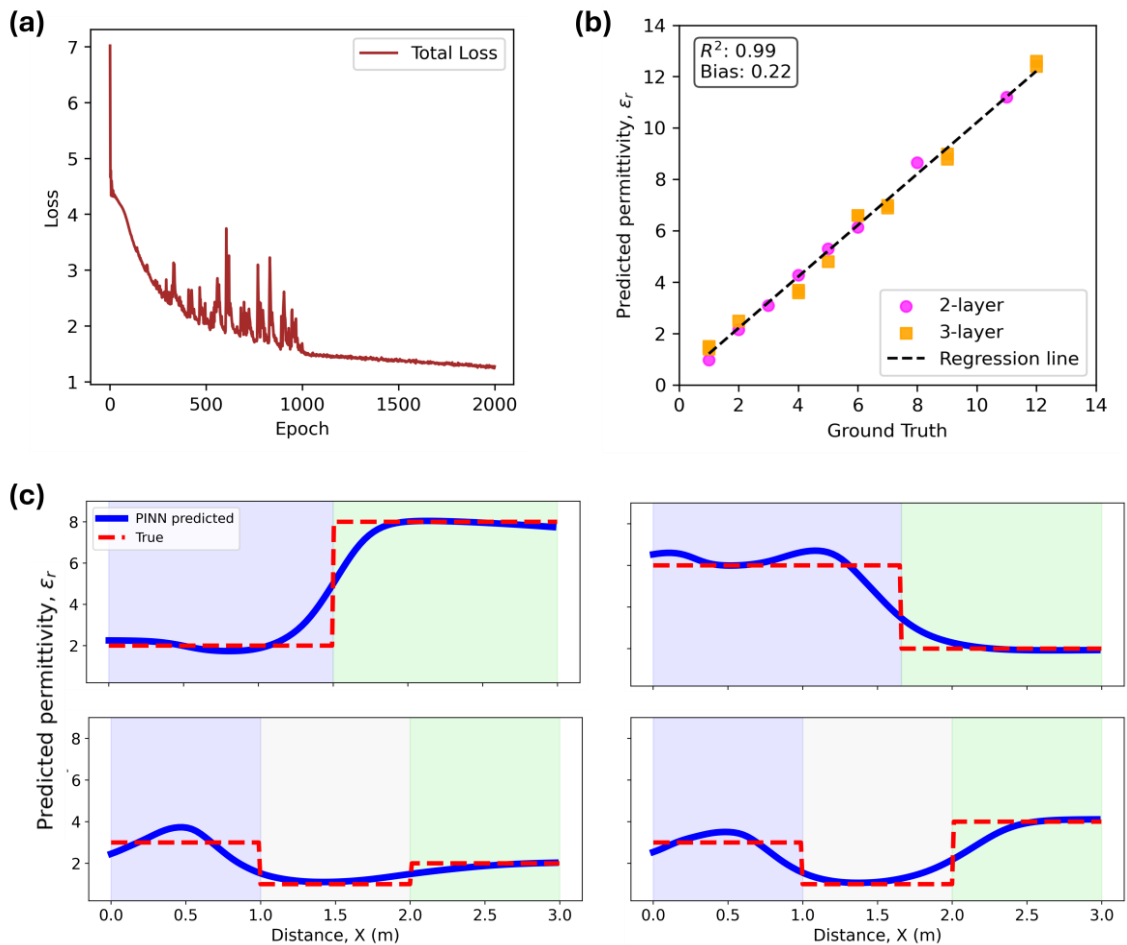

**Fig. 4: Performance of PINN on synthetic data;** (a) Total loss vs epoch during PINN training; (b) True vs. predicted permittivity of discrete known material layers; (c) Prediction of continuous permittivity profile.

## 2.3. Validation on real-world experiments

### 2.3.1. Radar system

To implement the proposed method in real-world experiments as per Fig. 3, an EM wave source and an array of spatially distributed receivers are required. Most commercially available ground penetrating radar (GPR) systems contain one transmitting and one receiving antenna within a single, inseparable unit (Fig. 5-a), thereby precluding the spatial distribution of multiple receivers. Hence, to allow one transmitting source and multiple

Page 6 of 22

receiver sensors, we built a custom radar setup using a Vector Network Analyzer (VNA) and several radiofrequency (RF) antennas, as shown in Fig. 5(b). We used LibreVNA [32] and Log-periodic antennas in this study. The frequency range of LibreVNA is 100 kHz to 6 GHz, and that of the Log-periodic antennas is 0.85 to 6.5 GHz. Considering the attenuation characteristics and thickness of the material layers, we used a frequency band of 2-4 GHz to generate EM waves, providing a balance between penetration distance and resolution. The signal is transmitted via a designated transmitting antenna (Tx), while the remaining antennas (Rx-1 to Rx-n) function as receivers. The receiving antennas measure the transmitted signal in terms of the $S_{21}$ scattering parameters (Supplementary Figure 1-a,b), which characterize the transmission response from the source to each receiver. The measured $S_{21}$ parameters in the frequency domain were converted to time-domain signals using an inverse Fourier transform, iFFT (Supplementary Figure 1-c).

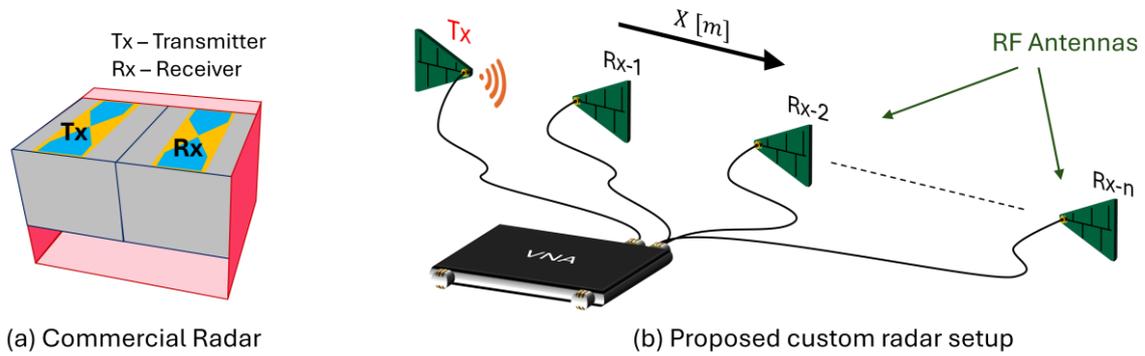

(a) Commercial Radar        (b) Proposed custom radar setup

**Fig. 5: The proposed custom radar system compared to a commercial radar.** (a) A typical commercial radar unit with integrated transmitter (Tx) and receiver (Rx) antennas; (b) The proposed custom radar system, which uses a vector network analyzer (VNA) connected to external radiofrequency (RF) antennas, enabling flexible multi-antenna configurations.

### 2.3.2. Materials

For real experiments, we created material layers using four different materials, viz. wood shavings, straw, soil, and air (Fig. 6-a). These materials were chosen because their actual permittivity can be readily measured using in-situ sensors, allowing direct comparison with the predicted permittivity. Moreover, wood shavings and straw closely represent the vegetation or biomass layer overlaying soil in a typical forest setup. We conducted an array of experiments by varying the materials, moisture contents, and the number of material layers, as shown in Table 1. The total length of material laid along the space is about 0.9 m for one-layer, 1.8 m for two-layer, and 2.7 m for three-layer experiments.

**Table 1: Matrix of experiments**

| Test # | No. of layers | Material(s) | Total length (m) |
|---|---|---|---|
| 1. | 1 | Air | 0.90 |
| 2. | 1 | Wood Shavings | 0.90 |
| 3. | 1 | Wet Sand | 0.90 |
| 4. | 2 | Straw + Dry Sand | 1.80 |
| 5. | 2 | Straw + Wet Sand | 1.80 |
| 6. | 2 | Wood Shavings + Straw | 1.80 |
| 7. | 3 | Wood Shavings + Straw + Dry Sand | 2.70 |

| 8. | 3 | Wood Shavings + Straw + Wet Sand | 2.70 |
| 9. | 3 | Air + Wood Shavings + Straw | 2.70 |

### 2.3.1. Experimental setup, data collection, and evaluation

The material setup was arranged based on Table 1. As an example, Fig. 6(c) shows the experimental setup and schematic view for a two-layer configuration with air and wood shavings, where the RF antennas (denoted by Rx) are spatially distributed for data collection. Fig. 6(d & e) also shows the schematic views of a typical one-layer and a three-layer experimental setup.

Due to the wide radiation pattern of the log-periodic antennas and the attenuation of signals within materials, the transmitted signal gradually decreases in amplitude as it propagates over distance. To maintain adequate signal strength across the sensing domain, we employed multiple transmitting positions (e.g., Tx-1, Tx-2, and Tx-3) by placing Tx antennas typically after every three Rx antennas. This setup is both practical and scalable, since in real-world applications, the number and arrangement of sensors can be adjusted as needed to ensure sufficient spatial coverage. For instance, receivers Rx-1 to Rx-4 receive the signal transmitted by Tx-1, while Rx-4 to Rx-7 capture the signal transmitted by Tx-2, and so on (see Fig. 6-e, for example). Notably, Rx-4 acts as the overlapping receiver between transmitter Tx-1 and Tx-2, and Rx-7, between Tx-2 and Tx-3. All received signals are first normalized and converted to absolute values to ensure consistency in magnitude and phase interpretation (Supplementary Figure 1-d & e). These preprocessed signals are then concatenated in the time domain using the overlapping receivers (e.g., Rx-4 and Rx-7) as reference points. Finally, we used the resulting composite dataset, which integrates measurements from all receivers, to train the PINN model.

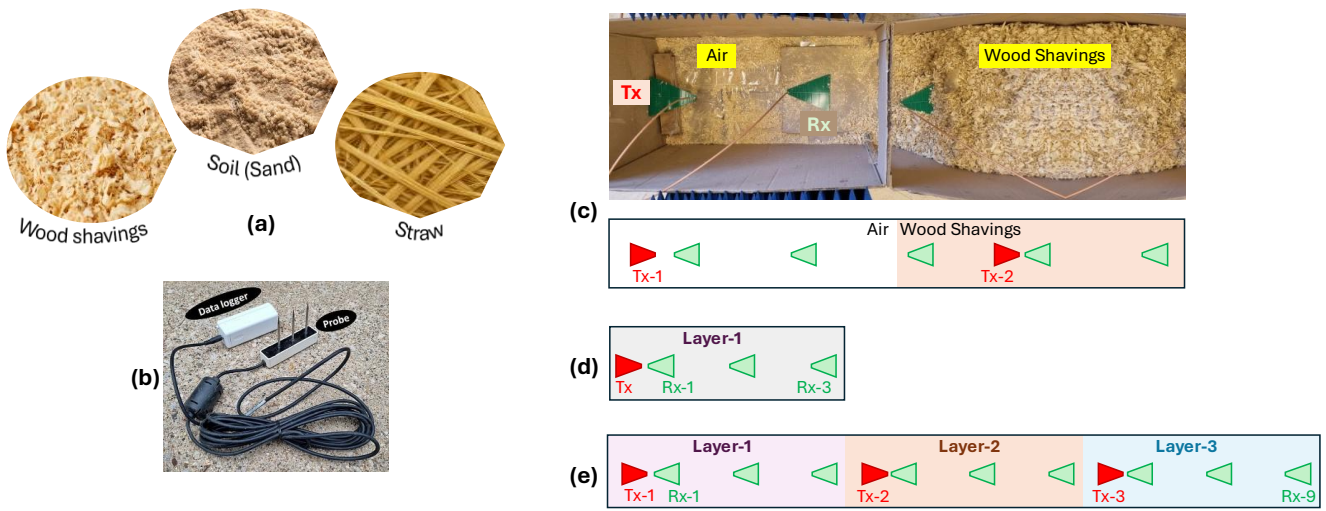

**Fig. 6: Materials, experimental setup, and evaluation.** (a) Materials used in the study; (b) In-situ sensor to measure permittivity; (c) Two-layer experiment with air and wood shavings and corresponding schematic view; (d) Schematic of one-layer configuration; (e) Schematic of three-layer configuration

For evaluating PINN's predictions, we measured the actual permittivity of each material approximately every 20 cm using an in-situ capacitance sensor [14], shown in Fig. 6(b). The permittivity values predicted by PINN were then compared with the in-situ sensor measurements to evaluate PINN's prediction accuracy.

### 2.3.2. Performance in the prediction of permittivity

First, the number of layers and the spatial location of the layer boundaries are assumed to be known. Under this assumption, we applied PINN model-1 of Fig. 2(d) to retrieve the permittivity of the discrete layers in the experiment. The permittivity values retrieved by PINN for all cases in Table 1 are compared against the in-situ sensor measurements in Fig. 7, demonstrating strong agreement between the predicted and measured values.

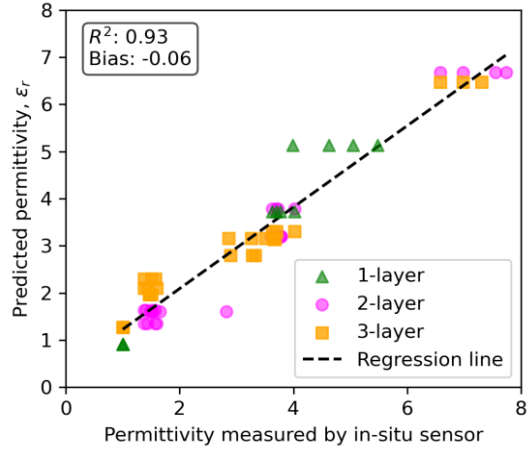

**Fig. 7: True vs. predicted permittivity of discrete known material layers – performance on experimental data**

Next, the number of layers and the spatial location of the layer boundaries were assumed to be unknown. Accordingly, we used PINN model-2 of Fig. 2(e) to retrieve the continuous permittivity profiles. Supplementary Movie 1 illustrates the loss and permittivity prediction over the epochs of PINN training.

To ensure repeatability and assess model robustness, we executed the PINN-training-prediction cycle 20 times with different weight initializations. For each test scenario, Fig. 8 presents the mean prediction across the 20 runs along with the ±1 standard deviation. In-situ sensor measurements recorded roughly every 20 cm are also shown in the same plots for comparison. The predicted permittivity profiles align well with the in-situ measurements, thereby validating the accuracy of the proposed method. The PINN framework successfully captures variations in permittivity for one-material (Fig. 8 a-c), two-material (Fig. 8 d-f), and three-material (Fig. 8 g-i) cases. Notably, the model is sensitive enough to detect even subtle changes in permittivity, such as a difference of $\varepsilon_r \approx 2$ in Fig. 8(f).

Traditional wave inversion methods typically require prior model calibration to accurately replicate real radar response and excitation, before executing the inversion algorithm or model-updating cycle [1][3][4][20][21][33][38]. In contrast, our proposed framework eliminates this requirement, as the transmitted radar pulse is implicitly learned by the PINN model directly from the measured data.

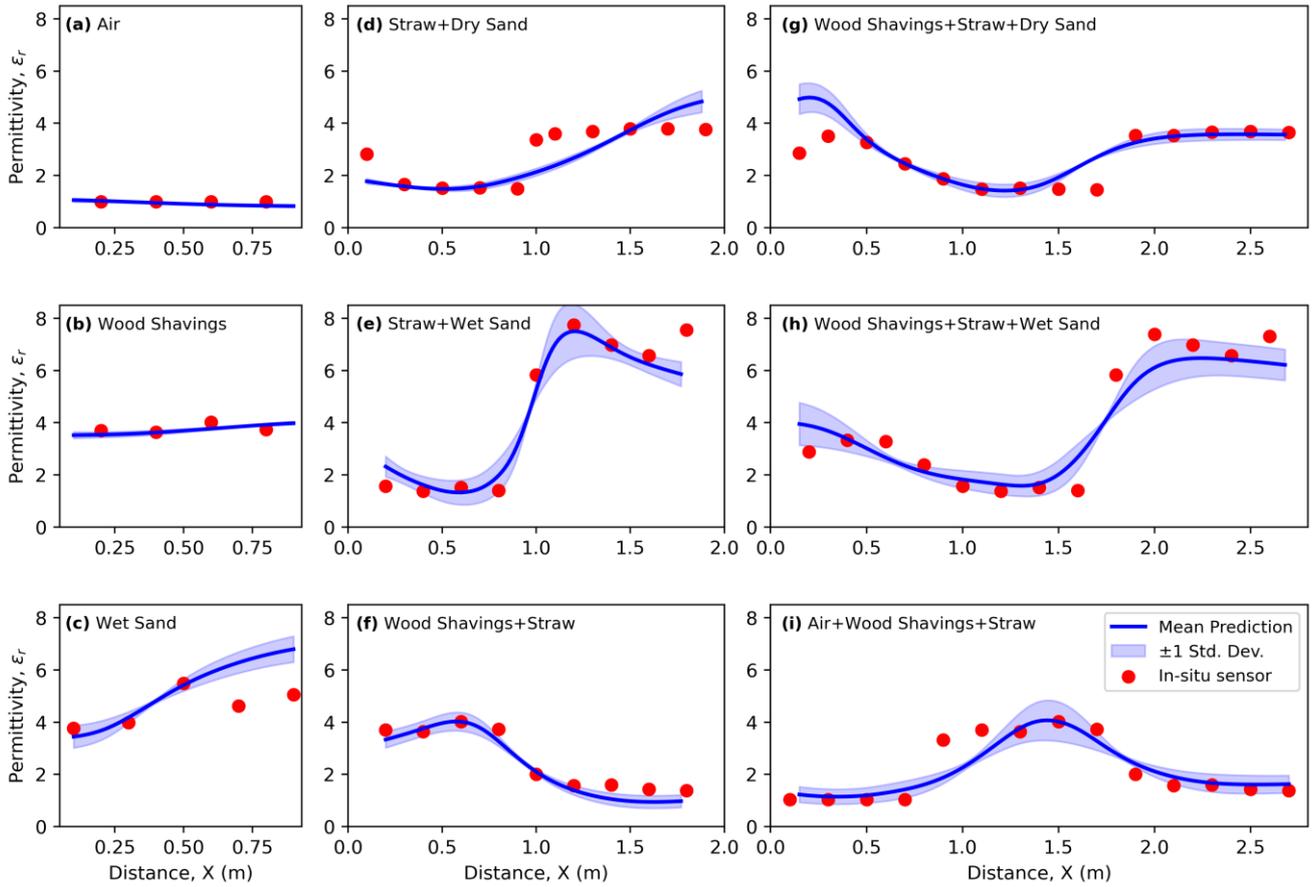

**Fig. 8: Continuous permittivity profiles predicted by PINN.** (a) to (i) represent the prediction results for the nine test cases listed in Table 1.

### 2.4. Accuracy vs. number of Rx-sensors

To determine the minimum number of receivers (Rx-sensors) required to accurately retrieve the spatial distribution of permittivity, we conducted a parametric analysis on the two-layer system, 'Wood Shavings + Straw'. As shown in Fig. 9(a), six candidate Rx-sensor positions were defined across the sensing domain, each recording waveform data to train the PINN and reconstruct the permittivity profile. We assessed the predictive accuracy of the retrieved profile using the mean squared error (MSE) between the PINN-predicted permittivity values and those measured by in-situ sensors at nine discrete spatial locations.

Fig. 9(b) shows the MSEs for different numbers of Rx used. When all six Rx-data were used, the prediction MSE was found to be very low. Next, we evaluated cases with five sensors: all six possible 5-Rx combinations (Rx-1-2-3-4-5, Rx-1-2-3-4-6, Rx-1-2-3-5-6, Rx-1-2-4-5-6, Rx-1-3-4-5-6, and Rx-2-3-4-5-6) were evaluated independently to train PINN and retrieve permittivity profiles. The average MSE across all combinations is shown in Fig. 9(b), along with its standard deviation. This procedure was repeated for combinations of 4, 3, and 2 Rx-sensors. As expected, prediction error consistently decreases as the number of Rx-sensors increases, indicating improved accuracy.

The retrieved profiles for all the different combinations of 2 to 6 Rx are shown in Supplementary Figure 2, while representative cases are illustrated in Fig. 9(c-h). For this two-material scenario, the permittivity profile could be closely reconstructed using all six sensors and any combination of five. Acceptable accuracy was also achieved with certain three- or four-sensor configurations—specifically, when one sensor was placed near the material boundary and the others were distributed within each layer (Fig. 9 – d, e, g, & h). In contrast, when all three or four Rx sensors were confined within the same material layer (Fig. 9 - c & f), predictions deteriorated due to the absence of waveform information from the other layer. Hence, the accuracy for three- or four-sensor cases varies greatly depending on sensor placement, which is reflected in the larger MSE variability. With only two Rx-sensors, the PINN fails to capture the permittivity profile across the domain, regardless of where they are placed, leading to the highest observed MSE.

Based on this analysis, evidence suggests that, if prior information about the material boundary is available, satisfactory estimation of the permittivity profile in a two-layer system can be achieved with as few as three strategically placed Rx-sensors. Similar parametric analysis can be extended to systems with three or more layers to identify optimal configurations. The two-layer case shown here illustrates the minimum Rx-sensor count and placement needed to resolve a sharp permittivity contrast across a material boundary. Without prior knowledge of material boundaries, we recommend placing Rx-sensors at equal intervals across the domain to ensure coverage of all possible regions of variation.

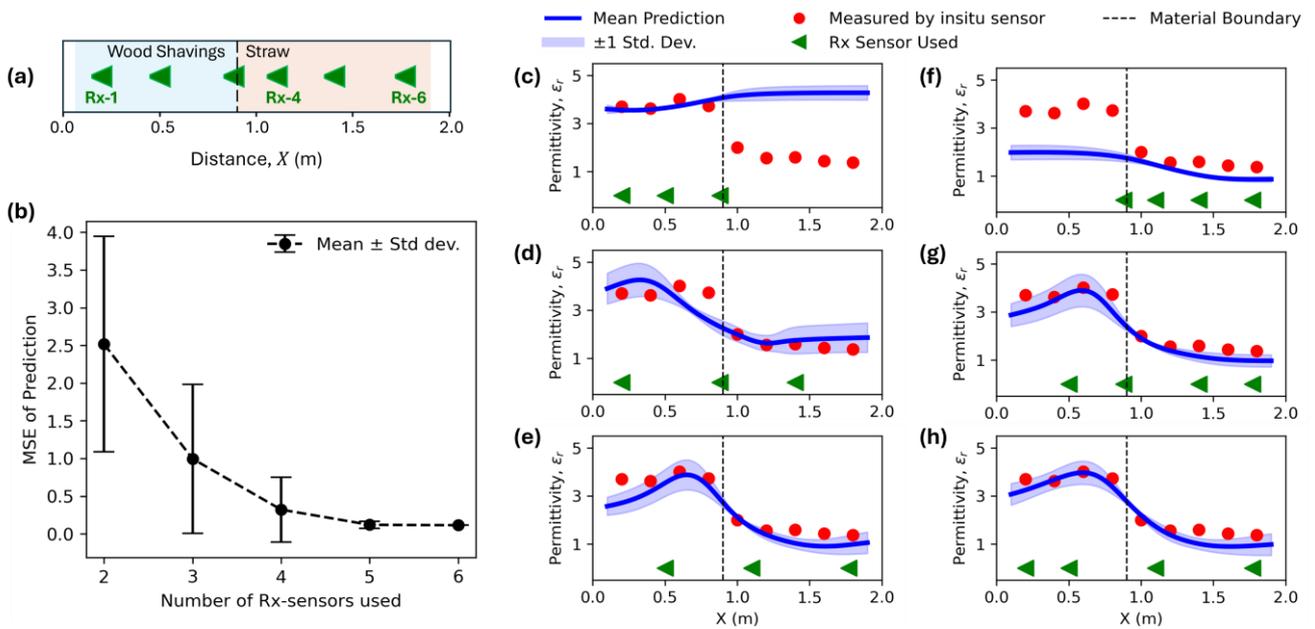

**Fig. 9: Study on accuracy vs. number of Rx-sensors used.** (a) Six possible locations of Rx-antennas considered in two-material layers; (b) Mean-squared error of prediction for different number of sensor data used to train PINN; (c, d & e) Shows retrieved permittivity profiles using 3 Rx; (f, g, & h) Shows retrieved permittivity profiles using 4 Rx. Rx-sensors used are shown as green triangles at the bottom of the same plots (c to h).

## 2.5. Outlook and limitations

We proposed a PINN-based framework to retrieve continuous subsurface property profiles, such as permittivity, using sparse radar measurements. By enforcing Maxwell's equations as physical constraints during neural network training, the model learns a fully neural representation of permittivity as a function of depth. This enables accurate recovery of smooth spatial variations with limited waveform data (as few as 3 to 6 measurement points). The framework performs effectively both with and without prior knowledge of subsurface layer counts or boundary locations, eliminating the need for discretized layer assumptions.

To support this, we developed a custom radar sensing system using a vector network analyzer (VNA) and distributed RF antennas. This design allows flexible configuration of frequency band and spatial sampling. This stands in contrast to typical commercial ground penetrating radar systems, which generally operate at a fixed center frequency and rely on rigid co-located transmitter-receiver units. Moreover, sensing data from multiple receivers increases the degree of information (DOI) [31], thereby improving inversion stability and mitigating non-uniqueness.

Despite these advances, some limitations remain. As shown in Fig. 5(b), the RF antennas are connected to the VNA using coaxial cables, which may be cumbersome or impractical for deployment across very long profiles. Radiofrequency reflective tags [37] can be used to enable passive wireless sensing. The tags reflect incoming radar signals and could emulate the spatial diversity achieved here without requiring physical wiring. Additionally, long-term deployment of buried antennas can be labor–intensive in soft materials such as soil, and infeasible in hardened materials such as concrete. Special protective casings with application-specific characteristics, such as dielectric transparency, biodegradability, or long-term durability, will need to be designed.

Overall, this study represents a significant step toward continuous, boundary-free subsurface profiling using nondestructive radar sensing. The framework's flexibility and data efficiency make it broadly applicable across domains. Beyond soil and vegetation moisture for agriculture and wildfire risk assessment, the predicted dielectric properties can support diverse applications, including active-layer permafrost characterization [5], density profiling of multilayered asphalt pavements [16], and moisture and corrosion mapping in concrete structures [15][39].

## 3. Methods

### 3.1. Physics-Informed Neural Network (PINN)

In a physics-informed neural network (PINN) model, the physics of a problem, i.e., the governing partial differential equations, are integrated into the loss function of the neural network training [22][23][25]. The physics (governing equations) of EM wave propagation is expressed by first-order coupled Maxwell's equations, provided in Eq. 1 and Eq. 2. **E** and **H** denotes the electric field and the magnetic field; $\mu, \varepsilon$ and $\sigma$ respectively denotes magnetic permeability, electric permittivity, and electrical conductivity of the medium in which the wave is propagating. These equations can also be represented in terms of a single field variable (**E**) by applying curl to Eq. 1. The derivation is shown in Supplementary Appendix A, and the derived form is Eq. 3, which is the scalar 2nd-order wave equation. This simplified form ignores conductivity ($\sigma = 0$) of material media. This assumption is reasonable, as a change in conductivity only changes the signal amplitude [33], and amplitude variation is mitigated here through normalization during data preprocessing (Supplementary Figure 1-c,d). Additionally, subsurface moisture—a key property of interest—is more strongly correlated with permittivity [1][15][20], which is already included as a variable in Eq. 3.

$$\nabla \times \mathbf{E} = -\mu \frac{\partial \mathbf{H}}{\partial t} \qquad \text{Eq. 1}$$

$$\nabla \times \mathbf{H} = \sigma \mathbf{E} + \varepsilon \frac{\partial \mathbf{E}}{\partial t} \qquad \text{Eq. 2}$$

$$\frac{\partial^2 E}{\partial t^2} - \frac{1}{\mu \varepsilon} \frac{\partial^2 E}{\partial x^2} = 0 \qquad \text{Eq. 3}$$

As shown in Fig. 2, space *x* and time *t* are inputs to the neural networks. Since *x* is expressed in meters and *t* is in nanoseconds ($\sim 10^{-9}$ seconds), the time values are multiplied by $10^9$ (before being input to the networks) to bring both *x* and *t* into a similar numerical scale. This transformation helps avoid gradient imbalance and improve numerical stability during network training. To account for this transformation, a scaling factor, $C = 10^{-18}$, is incorporated in Eq. 3 and the modified equation is shown in Eq. 4.

$$\frac{\partial^2 E}{\partial t^2} - \frac{1}{\mu \varepsilon} \frac{\partial^2 E}{\partial x^2} \times C = 0 \qquad \textbf{Eq. 4}$$

The objective of PINN training is to minimize the weighted sum of the data loss ($\mathcal{L}_{data}$) and the physics loss ($\mathcal{L}_{PDE}$), shown in Eq. 5:

$$Total\ loss,\ \mathcal{L} = \lambda_{data} . \mathcal{L}_{data} + \lambda_{PDE} . \mathcal{L}_{PDE} \qquad \textbf{Eq. 5}$$

where,

$$\mathcal{L}_{data} = \frac{1}{N_{data}} \sum_{i=1}^{N_{data}} (E_i - E_i^*)^2$$

$$\mathcal{L}_{PDE} = \frac{1}{N_{PDE}} \sum_{j=1}^{N_{PDE}} \left( \frac{\partial^2 E_j}{\partial t_j^2} - \frac{1}{\mu \varepsilon_j} \frac{\partial^2 E_j}{\partial x_j^2} \times C \right)^2$$

where $\lambda_{data}$ and $\lambda_{PDE}$ are scalar weights that balance the contribution of each loss term. The data loss is computed as the mean squared error (MSE) between the wavefield (*E*) predicted by the field network and the wavefield recorded by the receiver sensors (*E\**) at $N_{data}$ number of points.

The physics loss term, $\mathcal{L}_{PDE}$ ensures that the predicted electric field *E(x,t)* adheres to the governing PDE. To estimate this, first, $N_{PDE}$ number of spatial and temporal points are randomly generated within the full range of the spatial and temporal domains. These points, generated to assess the physics loss, are also called collocation points. The field network is then used to predict the electric field at those collocation points. The second-order derivatives of the predicted field with respect to space and time are computed using automatic differentiation [34]. Permittivity is predicted at spatial collocation points using the permittivity network, which outputs the relative permittivity $\varepsilon_{r_j}$. The absolute permittivity is then obtained as $\varepsilon_j = \varepsilon_0 . \varepsilon_{r_j}$, where $\varepsilon_0 = 8.854 \times 10^{-12}$ is the permittivity of vacuum. Next, at each collocation point *j*, the residual of the wave equation $\left| \frac{\partial^2 E_j}{\partial t_j^2} - \frac{1}{\mu \varepsilon_j} \frac{\partial^2 E_j}{\partial x_j^2} \times C \right|$ is computed (magnetic permeability $\mu = \mu_0 = 4\pi \times 10^{-7}$, since the materials used are non-magnetic). The physics loss is then estimated as the mean-squared residual across all collocation points. Minimization of this residual loss guides the model not only by the observed data but also by the governing physical law, allowing for better generalization in regions lacking direct observations.

At every epoch of PINN training, the network parameters are updated to minimize both the data loss and the physics loss (Eq. 5). We used the Adam optimizer with an initial learning rate of 0.01. To facilitate stable convergence, we also used a learning rate scheduler that decays the learning rate by a factor of 0.1 every 1000 training steps. Each network in Fig. 2 (d & e) has 2 hidden layers and 64 neurons in each hidden layer. At the hidden layers, we used Tanh as an activation function, and at the output of the permittivity network, we used an exponential activation function to enforce the positivity of permittivity.

## 3.2. Generation of synthetic data using FDTD

We used the finite-difference-time-domain (FDTD) [35][36] method to generate synthetic data. It can simulate EM wave propagation by solving either the coupled first-order equations (Eq. 1 and Eq. 2) or only the scalar wave equation (Eq. 3). Details of the FDTD method of radar wave propagation can be found in one of the authors' publications — Aziz & Alipour, 2025 [33].

In our FDTD simulations, the 1D EM wave transmitted by the source is modeled as a combination of a Sine and a Gaussian function, as shown in Eq. 6. *S(k)* represents the source signal at time step *k*, A is the amplitude (=2), *f* is the frequency (=0.5 GHz), and Δt is the time step (= $1.991 \times 10^{-11}$ s). The parameters *a* (=20) and *b* (=8) define the shape of the Gaussian envelope.

$$S(k) = 20 \cdot sin(A\pi f \cdot \Delta t \cdot k) \cdot e^{-0.5 \cdot \left(\frac{k-a}{b}\right)^2}$$   **Eq. 6**

Spatially distributed Rx-sensors (along distance in Fig. 3) record EM waveforms. The recorded waveforms are used as observed data to train the PINN model and retrieve either the discrete permittivity values (using the model of Fig. 2-d) or continuous permittivity profile (using the model of Fig. 2-e) along space (or distance).

## Acknowledgments

This work was supported by the NASA FireSense Technology Program grant 121509. The authors also acknowledge Keysight Technologies for additional support and NVIDIA Corporation's donation of GPUs used in this study. We further thank Dr. Elahe Soltanaghai for technical feedback, and Luke Jacobs and Avery Plote for their assistance in data collection.

## Data availability

The experimental data are available in this GitHub repository: PINN-Radar-Permittivity-Inversion/Exp_data at main · ishfaqa2/PINN-Radar-Permittivity-Inversion

## Code availability

The Python codes used in this study are made available in this GitHub repository: https://github.com/ishfaqa2/PINN-Radar-Permittivity-Inversion.git.

# Supplementary file

## Appendix A: Deriving 2nd order wave equation for EM waves

Taking curl of Eq. 1:

$$\nabla \times (\nabla \times \boldsymbol{E}) = \nabla \times \left(-\mu \times \frac{\partial \mathbf{H}}{\partialت}\right) \quad \text{Eq. 7}$$

Applying the vector identity on the left-hand side:

$$\nabla \times (\nabla \times \boldsymbol{E}) = \nabla(\nabla \cdot \mathrm{E}) - \nabla^2 \boldsymbol{E};$$

Since $\nabla \cdot \mathbf{E} = 0$ (no free charges), we simplify: $\quad \nabla \times \nabla \times \boldsymbol{E} = -\nabla^2 \boldsymbol{E}$

The Laplacian operator is given by $\nabla^2 = \frac{\partial^2}{\partial x^2} + \frac{\partial^2}{\partial y^2} + \frac{\partial^2}{\partial z^2}$

Substituting into Eq. 7:

$$\nabla^2 \boldsymbol{E} = \mu \frac{\partial}{\partial \mathrm{t}} (\nabla \times \mathbf{H})$$

$$\Rightarrow \nabla^2 \boldsymbol{E} = \mu \frac{\partial}{\partial \mathrm{t}} \left(\sigma \mathbf{E} + \varepsilon \frac{\partial \mathbf{E}}{\partial \mathrm{t}}\right) \quad \text{[Using Eq. 2]}$$

$$\Rightarrow \nabla^2 \boldsymbol{E} = \mu\sigma \frac{\partial \mathbf{E}}{\partial \mathrm{t}} + \varepsilon\mu \frac{\partial^2 \mathbf{E}}{\partial \mathrm{t}^2}$$

$$\Rightarrow \nabla^2 \boldsymbol{E} - \mu\sigma \frac{\partial \mathbf{E}}{\partial \mathrm{t}} - \mu\varepsilon \frac{\partial^2 \mathbf{E}}{\partial \mathrm{t}^2} = \mathbf{0} \quad \text{Eq. 8}$$

For media with $\sigma = 0$, and using the EM wave velocity: $c = 1/\sqrt{\mu\varepsilon}$, Eq. 8 reduces to

$$\nabla^2 \boldsymbol{E} - \frac{1}{c^2} \frac{\partial^2 \mathbf{E}}{\partial \mathrm{t}^2} = \mathbf{0} \quad \text{Eq. 9}$$

The 1D form of Eq. 9 is:

$$\frac{\partial^2 \boldsymbol{E}}{\partial t^2} - \frac{1}{\mu\boldsymbol{\varepsilon}} \frac{\partial^2 \boldsymbol{E}}{\partial x^2} = 0 \quad \text{Eq. 10}$$

# Appendix B: Figures

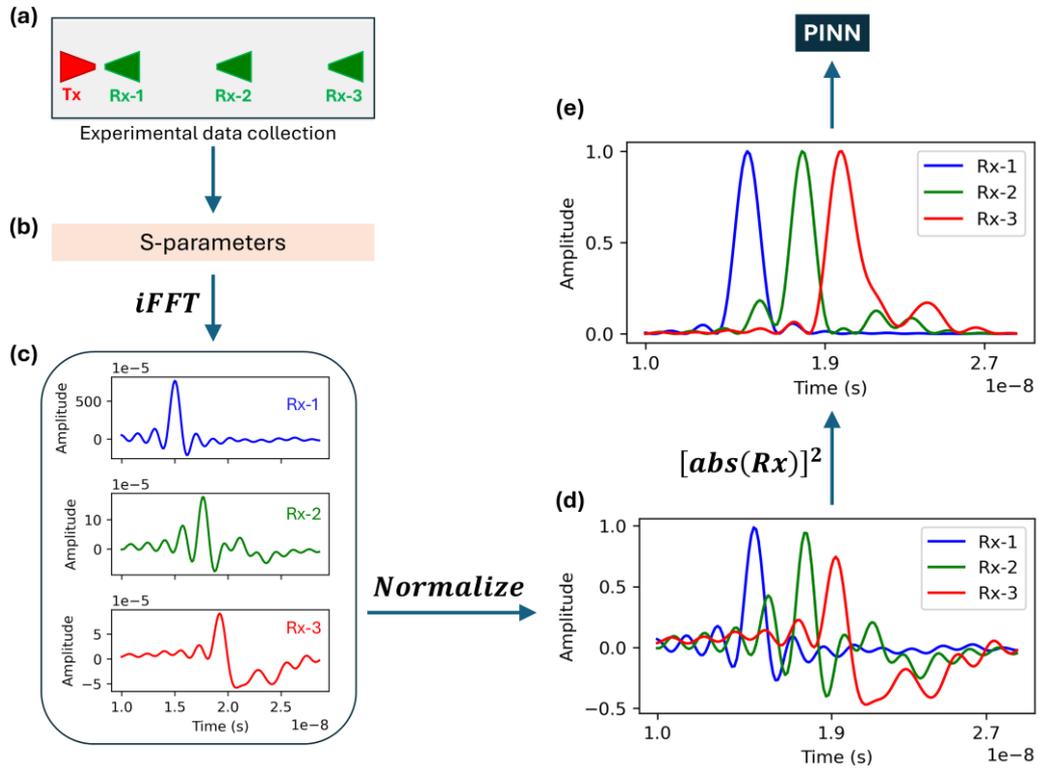

**Figure 1: Typical data processing steps.** (a) Data collection with VNA and RF antennas (denoted by Rx here); (b) S-parameters measured by VNA at spatial locations of Rx; (c) Time-domain signals at Rx-locations obtained after inverse Fast Fourier Transform (iFFT); (d) Normalized time-domain signals at Rx-locations; (e) Squared absolute value of time-domain signals at Rx-locations – this is the final form of data fed to PINN for training and prediction.

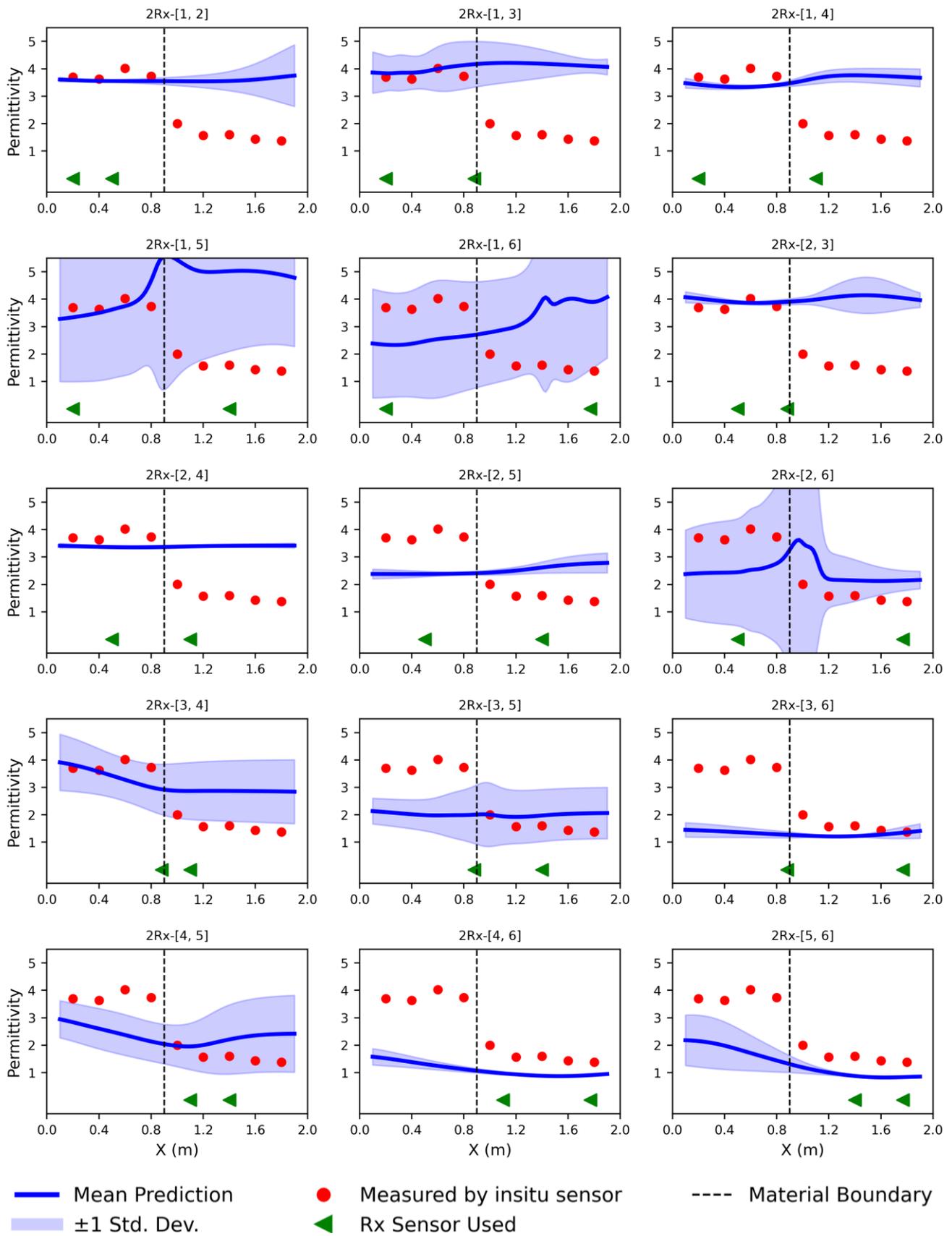

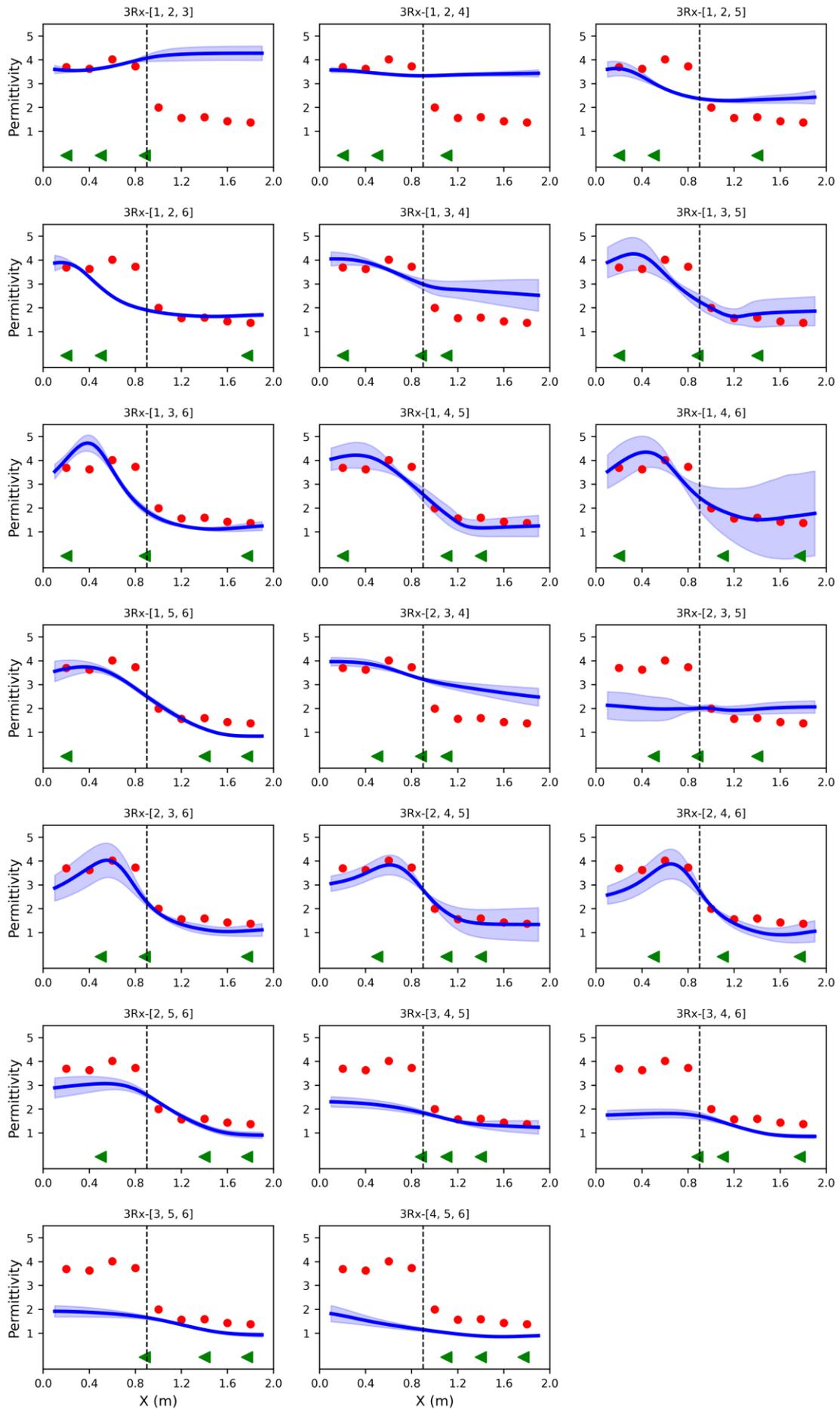

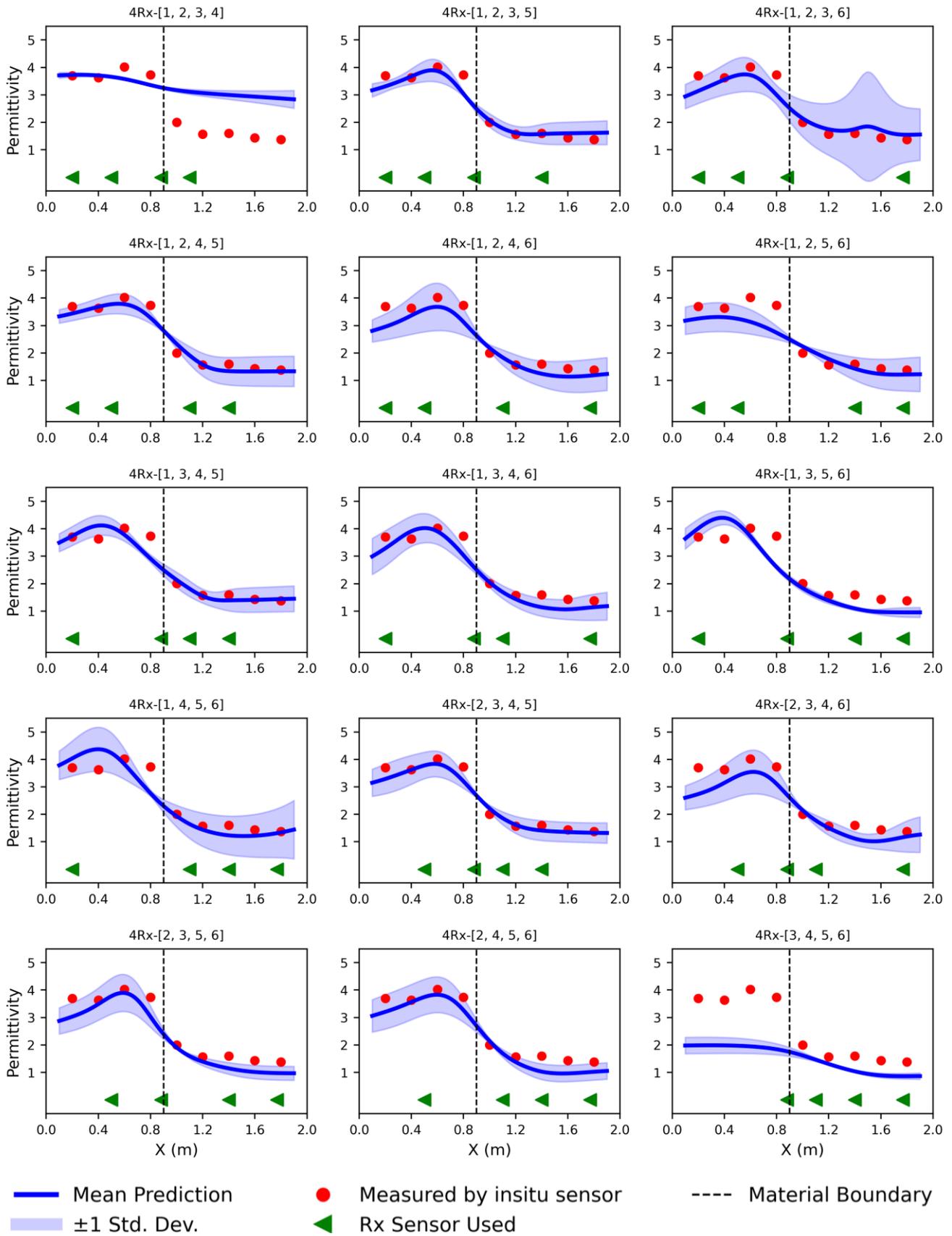

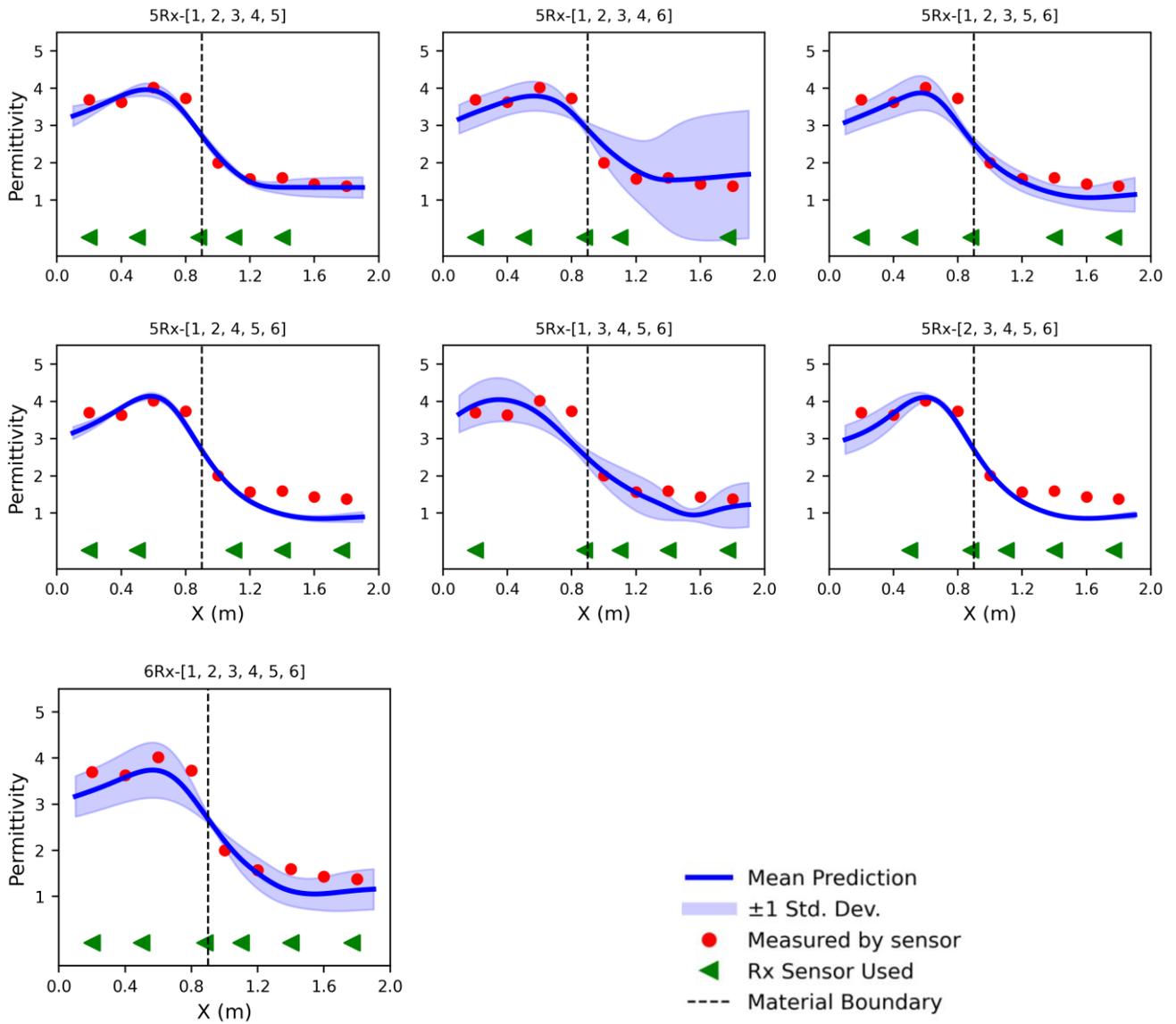

**Figure 2: PINN prediction for different numbers of Rx-antennas used**